\documentclass[%
 reprint,
 amsmath,amssymb,
 aps,
]{revtex4-2}

\usepackage{graphicx}
\usepackage{dcolumn}
\usepackage{bm}

\begin{document}

\preprint{APS/123-QED}

\title{Eigen states in the self-organised criticality}

\author{Yongwen Zhang}
\email{zhangyongwen77@gmail.com}
 \affiliation{%
 Data Science Research Center, Faculty of Science, Kunming University of Science
and Technology, Kunming 650500, China\\
}
\author{Maoxin Liu}%
\affiliation{%
School of Systems Science, Beijing Normal University, Beijing 100875, China\\
}%

\author{Gaoke Hu}
\affiliation{%
 School of Systems Science, Beijing Normal University, Beijing 100875, China\\
}%
\author{Teng Liu}
\affiliation{%
 School of Systems Science, Beijing Normal University, Beijing 100875, China\\
}%

\author{Xiaosong Chen}
\email{chenxs@bnu.edu.cn}
\affiliation{%
 School of Systems Science, Beijing Normal University, Beijing 100875, China\\
}%


\date{\today}

\begin{abstract}
We employ the eigen microstate approach to explore the self-organized criticality (SOC) in two celebrated sandpile models, namely, the BTW model and the Manna model. In both models, phase transitions from the absorbing-state to the critical state can be understood by the emergence of dominant eigen microstates with significantly increased weights. Spatial eigen microstates of avalanches can be uniformly characterized by a linear system size rescaling. The first temporal eigen microstates reveal scaling relations in both models. Furthermore, by finite-size scaling analysis of the first eigen microstate, we numerically estimate critical exponents i.e., $\sqrt{\sigma_0 w_1}/\tilde{v}_{1} \propto L^D$ and $\tilde{v}_{1} \propto L^{D(1-\tau_s)/2}$. Our findings could provide profound insights into eigen states of the universality and phase transition in non-equilibrium complex systems governed by self-organized criticality.

\end{abstract}

\maketitle


\section{\label{sec1}Introduction}
Phase transitions and critical phenomena play pivotal roles in understanding complex systems. Classical models, such as the Ising model and percolation model \cite{christensen_complexity_2005}, adeptly capture the continuous phase transition properties. In the Ising model, approaching a critical temperature leads to power-law behaviors in thermodynamic quantities like magnetization and susceptibility, with universal critical exponents which are independent microscopic details \cite{christensen_complexity_2005}. However, the applicability of the Ising model to understand critical phenomena in complex systems is constrained by its requirement for thermodynamic equilibrium, a condition seldom met by real complex systems, which predominantly exist in non-equilibrium states.

In 1987, Bak, Tang, and Wiesenfeld (BTW) established the concept of SOC by introducing the BTW sandpile model \cite{bak_self-organized_1987}. Within this model, power-law behaviors were identified in the size of avalanches, with potential implications for understanding phenomena in the real world. For instance, earthquake energy release follows a power-law distribution \cite{gutenberg1944frequency}, rainfall exhibits continuous phase transitions \cite{peters_critical_2006}, and the waiting time statistics of various natural records mirror the sandpile model's behavior \cite{bak_unified_2002,sanchez_waiting-time_2002,paczuski_interoccurrence_2005,deluca_data-driven_2015}.

The BTW model illustrates the spontaneous evolution of a non-equilibrium system toward a critical state marked by slow driving energy and rapid energy dissipation \cite{manna_sandpile_1999}. A continuous phase transition between an absorbing state and an active phase is a distinctive feature of the sandpile model \cite{dickman_self-organized_1998,vespignani_driving_1998,vespignani_absorbing-state_2000}. The BTW model, characterized by an Abelian group structure, draws an analogy with the Abelian Manna model, which incorporates random relaxation rules \cite{manna_two-state_1991,dhar_abelian_1999}. Zhang also proposed another SOC model involving continuous energy \cite{zhang_scaling_1989}. Experimental validation of SOC behavior was demonstrated through a rice pile experiment \cite{frette_avalanche_1996,christensen_tracer_1996}. The universality class of various SOC models, including both the BTW and Manna models, has been reported as identical \cite{pietronero_renormalization_1994,chessa_universality_1999}. Nevertheless, there exists some contention regarding the assertion that they might belong to distinct universality classes \cite{ben-hur_universality_1996,milshtein_universality_1998}.

In this paper, we introduce the concept of eigen microstates of the statistical ensemble to explore the critical behaviors of sandpile models. The critical state can be delineated by the condensed eigen microstate, characterized by the emergence of a substantial weight factor. This characterization finds support in the Ising model \cite{hu_condensation_2019,sun_eigen_2021,liu_renormalization_2022} and collective motion \cite{li_discontinuous_2021}. Our proposed methodology enables the identification of phase transitions and universality classes in non-equilibrium systems without requiring knowledge of the order parameter. 

In the subsequent section, we elucidate the definition of SOC sandpile models and articulate the concept of the eigen microstate within the statistical ensemble for the system. Section~\ref{sec3} provides a detailed exposition of the simulations conducted on SOC sandpile models, along with the presentation of results pertaining to their eigen microstates, analyzed through finite-size scaling. Conclusive remarks are then drawn in Section~\ref{sec4}.

\section{\label{sec2}MODEL DEFINITION AND METHODS}
\subsection{The SOC sandpile models}
We explore two models exemplifying SOC: the BTW model and the Abelian Manna model. For the BTW model, we examine a square lattice with $N=L^2$ sites, where $L$ represents the system's size. The initial state involves a random distribution of non-negative integer heights $z$ for sites, with a given average height. The system undergoes driving forces by adding a grain to a randomly chosen site $i$, resulting in an increase in height $z_i \rightarrow z_i+1$. When $z_i \geq z_0$ (where $z_0=4$ is a predefined threshold height), the site becomes unstable and topples (relaxes) as $z_i \rightarrow z_i-4$. Furthermore, each nearest neighboring site $j$ gains one grain, leading to $z_j \rightarrow z_j+1$. Toppling may induce instability in neighboring sites, triggering a cascade of toppling until stability is restored, with grains dissipating from the open boundary. This entire toppling process constitutes an avalanche. New grains are added for the next time step until the last avalanche concludes. Over multiple time steps, the system reaches the critical state characterized by recurrent configurations, identifiable through the Burning Test \cite{dhar_studying_1999}.

Contrastingly, the toppling rule differs in the Abelian Manna model, which introduces an element of randomness. In this model, if $z_i \geq 2$ at any site $i$, the site becomes unstable and topples, reducing $z_i$ to $z_i-2$. The two grains are then distributed randomly and independently among nearest neighbors, with the possibility of selecting the same site twice: $z_j \rightarrow z_j+1$. The remaining aspects of the Abelian Manna model mirror those of the BTW model described above.   

\subsection{Eigen microstates}
In the sandpile system with $N$ sites, we extract the states of the sites through simulations. By conducting measurements over $M$ time steps, we obtain an $N\times M$ matrix ($M>N$) representing the state as $\bm{S}=(\bm{s_1}, \bm{s_2}, \dots, \bm{s_t}, \dots, \bm{s_M})$, where $\bm{s_t}=(s_{1t}, s_{2t}, \dots, s_{Nt})^T$ signifies the microstate of the system at time step $t$, and $s_{it}$ is the number of topplings at site $i$ during time step $t$. This matrix $\bm{S}$ can be treated as a statistical ensemble with dynamic microstates of the sandpile system. Utilizing singular value decomposition (SVD), we decompose the ensemble matrix as $\bm{S}=\bm{U}\cdot\bm{\Lambda}\cdot\bm{V}^T$, where $\bm{\Lambda}$ is a $N\times N$ diagonal matrix with non-zero eigenvalues $\lambda_{1}\geq \lambda_{2}\dots\geq\lambda_{I}\dots\geq\lambda_{N}\geq 0$. The corresponding eigenvectors are represented by a $M\times N$ unitary matrix $\bm{V}=(\bm{v_1}, \bm{v_2}, \dots, \bm{v_I}, \dots, \bm{v_N})$ and a $N\times N$ unitary matrix $\bm{U}=(\bm{u_1}, \bm{u_2}, \dots, \bm{u_I}, \dots, \bm{u_N})$, where $\bm{v_I}=(v_{1I}, v_{2I}, \dots, v_{MI})^T$ and $\bm{u_I}=(u_{1I}, u_{2I}, \dots, u_{NI})^T$.

Furthermore, we express the ensemble matrix $\bm{S}$ as \cite{hu_condensation_2019}:
\begin{equation}\label{eq1}
\bm{S}=\sum_{I=1}^{N}w_I^{1/2}\bm{S_I^E}\;,
\end{equation}
where $\bm{S_I^E}$ is an $N\times M$ eigen ensemble matrix of the system with elements $(\bm{S_I^E})_{it}=C_0^{1/2}u_{iI}v_{tI}$. Here, the eigenvector $\bm{u_I}$ corresponds to the normalized eigen microstate, and the eigenvector $\bm{v_I}$ represents the normalized temporal eigen microstate. The constant amplitude is defined as $C_0=\sum_{I=1}^N\lambda_I^2=\sum_{t=1}^{M}\sum_{i=1}^{N}s_{it}^2$, and $w_I=\lambda_I^2/\sum_{I=1}^N\lambda_I^2$ serves as a weight associated with the probability of the eigen ensemble $\bm{S_I^E}$ in the statistical ensemble $\bm{S}$.

For a given time step $t$, the avalanche size is denoted as $A_s(t)=\sum_{i=1}^N s_{it}$. Consequently, the average avalanche size over $M$ steps is expressed as:
\begin{equation}\label{eq2}
\langle A_s \rangle=\sigma_{0}^{1/2}\sum_{I=1}^Nw_I^{1/2}\tilde{u}_{I}\tilde{v}_{I}\;,
\end{equation}
where $\sigma_0=\frac{NC_0}{M}$, and the parameters $\tilde{u}_{I}$ and $\tilde{v}_{I}$ are defined as:
\begin{eqnarray}
\tilde{u}_{I}=\frac{1}{\sqrt{N}}\sum_{i=1}^N u_{iI}\;,
\label{eq3:1}
\\
\tilde{v}_{I}=\frac{1}{\sqrt{M}}\sum_{t=1}^M v_{tI}\;.
\label{eq3:2}
\end{eqnarray}
Given $\sum_{t=1}^M v_{tI}^2=1$, the second moment of the avalanche size is calculated as:
\begin{equation}\label{eq4}
\langle A_{s}^{2} \rangle=\sigma_{0}\sum_{I=1}^N w_I\tilde{u}_I^2\;.
\end{equation}
In our study, we employ $L=16, 32, 64, 96, 128$ and $M=5\times10^{5}$ to calculate the eigen ensemble.

\section{\label{sec3}RESULTS}

\begin{figure}
\centering
\includegraphics[scale=0.3]{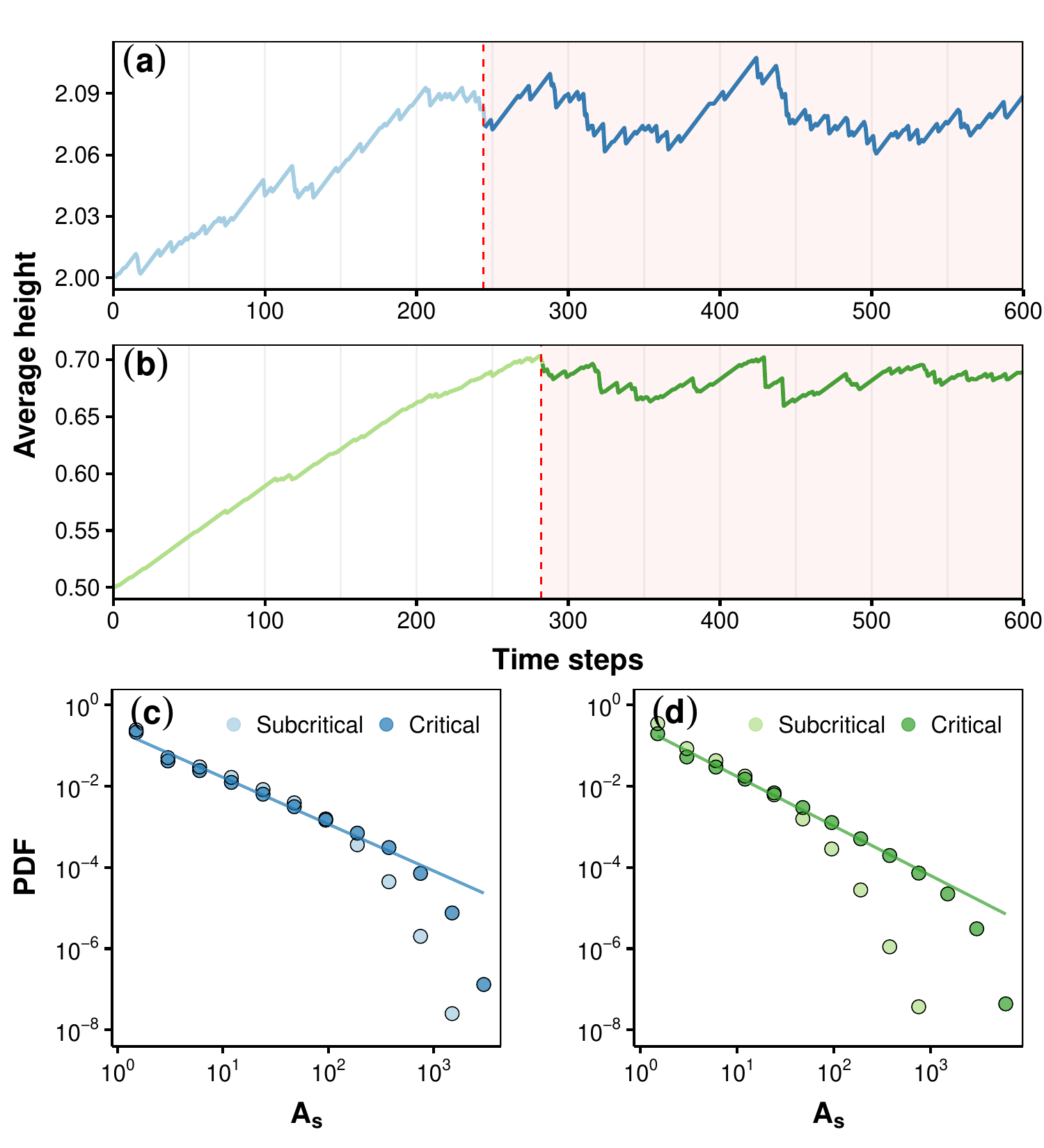}
\caption{\label{fig:1} Evolution of the average height of the sandpile from the absorbing state to the critical state for (a) the BTW model and (b) the Manna model with $L=32$. The system reaches the critical state after the time step indicated by the red dashed line. PDFs of the avalanche size in subcritical and critical states for (c) the BTW model and (d) the Manna model. The blue and green lines represent fitted lines with slopes of -1.1$\pm$0.07 and -1.17$\pm$0.04, respectively.}
\end{figure}

The simulations of the sandpile models, depicted in Fig.~\ref{fig:1}, illustrate the transition from the absorbing state to the critical state. In Fig.~\ref{fig:1}(a) and (b) (before the red dashed line), when the average height is low, the system resides in the absorbing state. Here, the average height increases proportionally with the addition of grains to the system. Subsequently, upon reaching the critical state, as evident in Fig.~\ref{fig:1}(a) and (b) (after the red dashed line), the average height stabilizes and fluctuates around a specific value. Throughout the critical state, the system's average energy remains constant for many time steps, indicative of a balance between energy input and dissipation. Both the BTW and Manna models exhibit a comparable evolution, albeit with distinct critical heights, as illustrated in Fig.~\ref{fig:1}(a) and (b).

An essential feature of SOC is its adherence to power-law behavior. The Probability Density Functions (PDFs) of avalanche sizes in subcritical and critical states are illustrated in Fig~\ref{fig:1}(c) and (d). In the subcritical state, the distribution of $A_s$ exhibits a faster decay as $A_s$ increases compared to the critical state. A proposed PDF satisfies $P(A_s)=A_s^{-\tau_s} G(A_s/A_{sc})$, where $\tau_s$ represents a critical exponent, and $G(x)$ is a scaling function. The cutoff for the system, denoted by $A_{sc}$ ($\sim L^D$), is determined by the system size, where $D=2.75$ is a critical exponent associated with the fractal dimension \cite{tebaldi_multifractal_1999}. Beyond the avalanche size $A_{sc}$, the PDF experiences a rapid decay. In Fig.~\ref{fig:1}(c) and (d), we determine the exponents $\tau_s=1.10\pm0.07$ and $1.17\pm0.04$ for the BTW model and the Manna model, respectively, with $L=32$. A correction factor ($\sim 1/\ln L$) can be applied to the critical exponent $\tau_s$. As $L$ approaches infinity, $\tau_s$ converges to 1.27 for both models \cite{chessa_universality_1999,lubeck_numerical_1997}.

\begin{figure}
\centering
\includegraphics[scale=0.3]{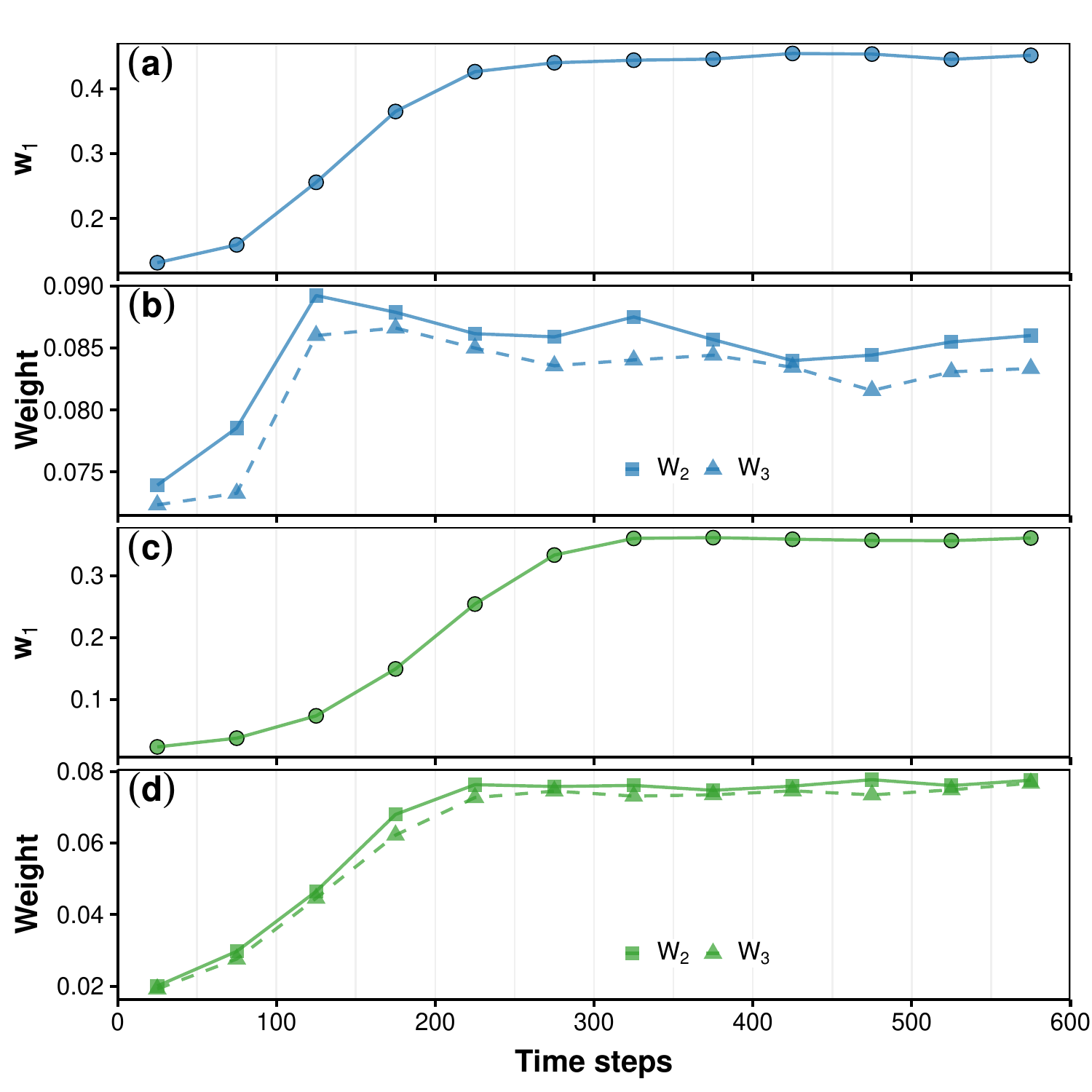}
\caption{\label{fig:2} Evolution of the weights: (a) $w_1$ and (b) $w_2$ and $w_3$ of the eigen ensemble from the absorbing state to the critical state for the BTW model with $L=32$. (c) and (d) Similar to (a) and (b), but for the Manna model. Each ensemble matrix is calculated based on $M=5\times10^{5}$ from independent $5\times10^{5}$ realizations.}
\end{figure}

We proceed to derive eigen ensembles and their associated weights according to Eq.~(\ref{eq1}). The evolution of these weights, from the absorbing state to the critical state, is presented in Fig.~\ref{fig:2}. Notably, for both the BTW and Manna models, the dominant eigen microstate emerges, linked with the amplified weight $w_1$, as shown in Fig.~\ref{fig:2} (a) and (c). Upon reaching the critical state, the substantial weight $w_1$ above 0.3 stabilizes, indicating the presence of a phase transition captured by the eigen microstate. It is akin to observations in the Ising model at the critical temperature \cite{hu_condensation_2019}. 

Additionally, for the smaller weights $w_2$ and $w_3$ (almost degenerate), we observe an incremental increase as the system evolves toward the critical state, as demonstrated in Fig.~\ref{fig:2}(c) and (d). However, it is noteworthy that the maximum values of the weights $w_2$ and $w_3$ emerge earlier than that of $w_1$. This early emergence could potentially serve as a valuable indicator for detecting early warning signals in SOC systems.

\begin{figure}
\centering
\includegraphics[scale=0.34]{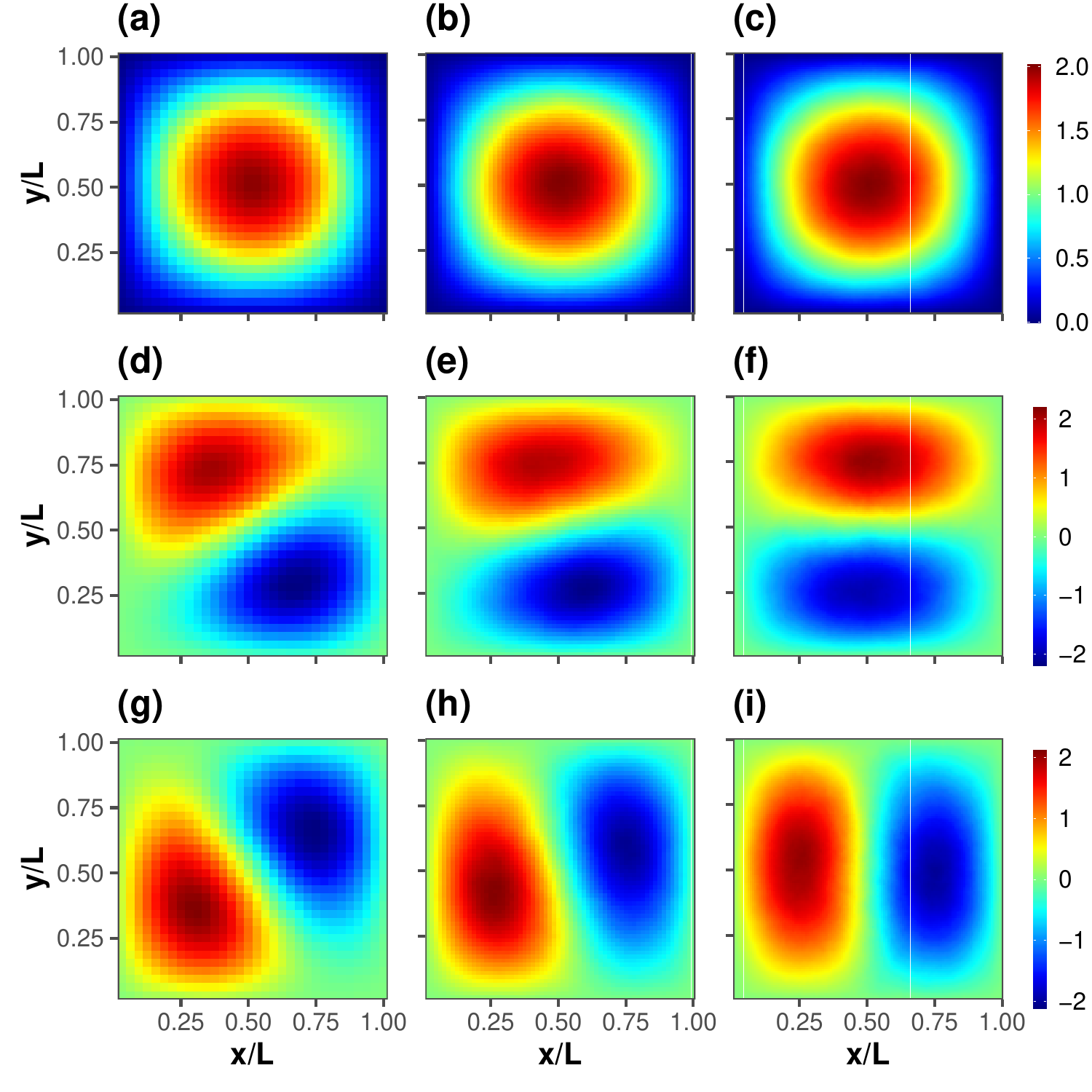}
\caption{\label{fig:3} Spatial distributions of the rescaled eigen microstate $\bm{u_1}L$ for the BTW model at the critical state with different system sizes: (a) $L=32$, (b) $L=64$, and (c) $L=128$. Correspondingly, (d-f) and (g-i) present the spatial distributions of $\bm{u_2}L$ and $\bm{u_3}L$ for the same system sizes, respectively.}
\end{figure}

Spatial distributions of the rescaled eigen microstates $\bm{u_1}L$, $\bm{u_2}L$, and $\bm{u_3}L$ at the critical state are illustrated in Fig.~\ref{fig:3} for the BTW model. The first eigen microstate reveals the universality and the presence of a giant cluster nearly matching the size of the system, with its components increasing from the boundary to the center of the system, as depicted in Fig.~\ref{fig:3}(a-c). 

\begin{figure}
\centering
\includegraphics[scale=0.34]{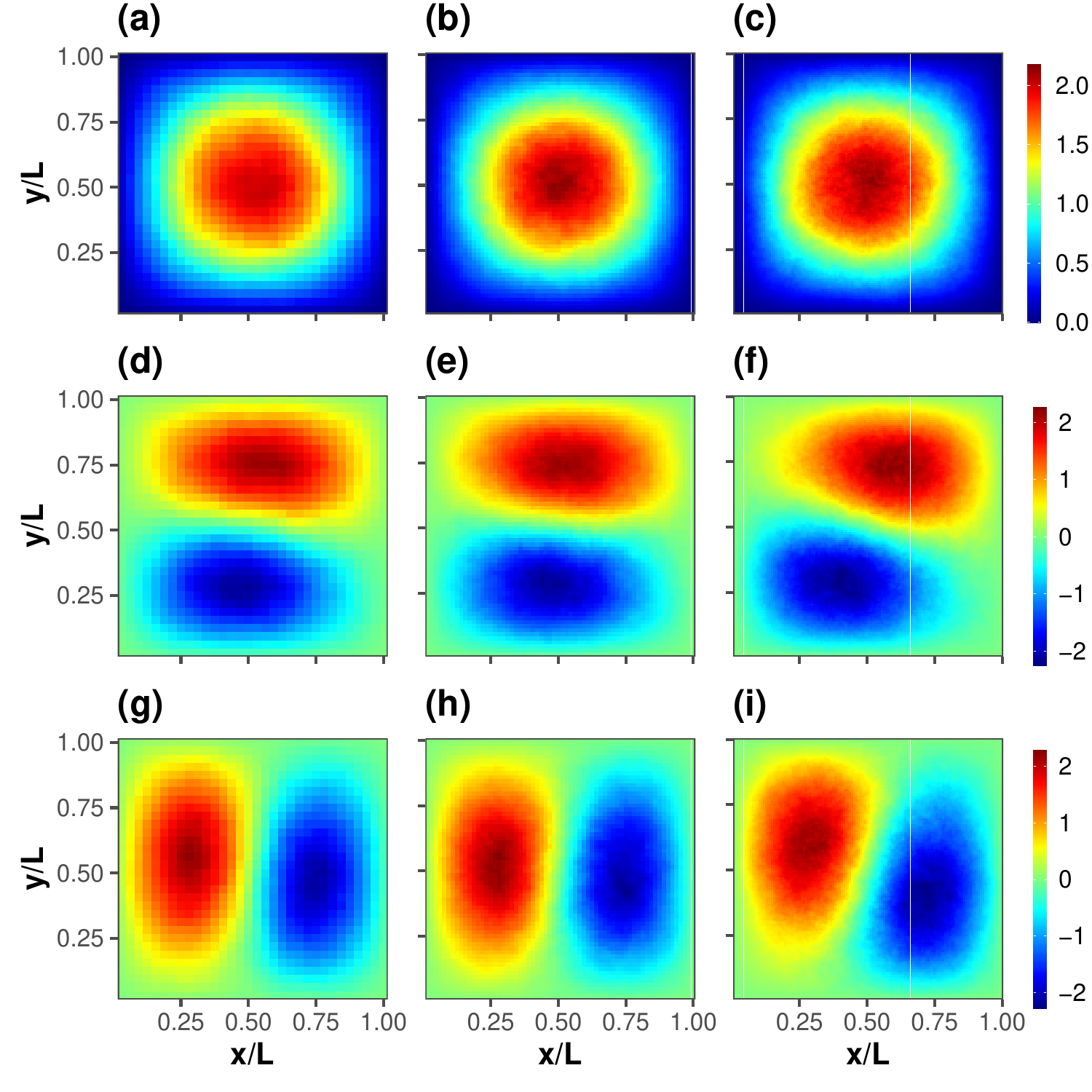}
\caption{\label{fig:4} Similar to Fig.~\ref{fig:3}, but for the Manna model.}
\end{figure}

The second-largest eigen microstate displays two clusters with opposite orientations, as seen in Fig.~\ref{fig:3}(d-f). Given that the eigenvectors are orthogonal, the third-largest eigen microstate will be rotated by $\pi/2$ relative to the second-largest eigen microstate, as shown in Fig.~\ref{fig:3}(g-i). These observed behaviors remain universality for different sizes. For the Manna model, the results closely mirror those of the BTW model in Fig.~\ref{fig:4}. 

\begin{figure}
\centering
\includegraphics[scale=0.34]{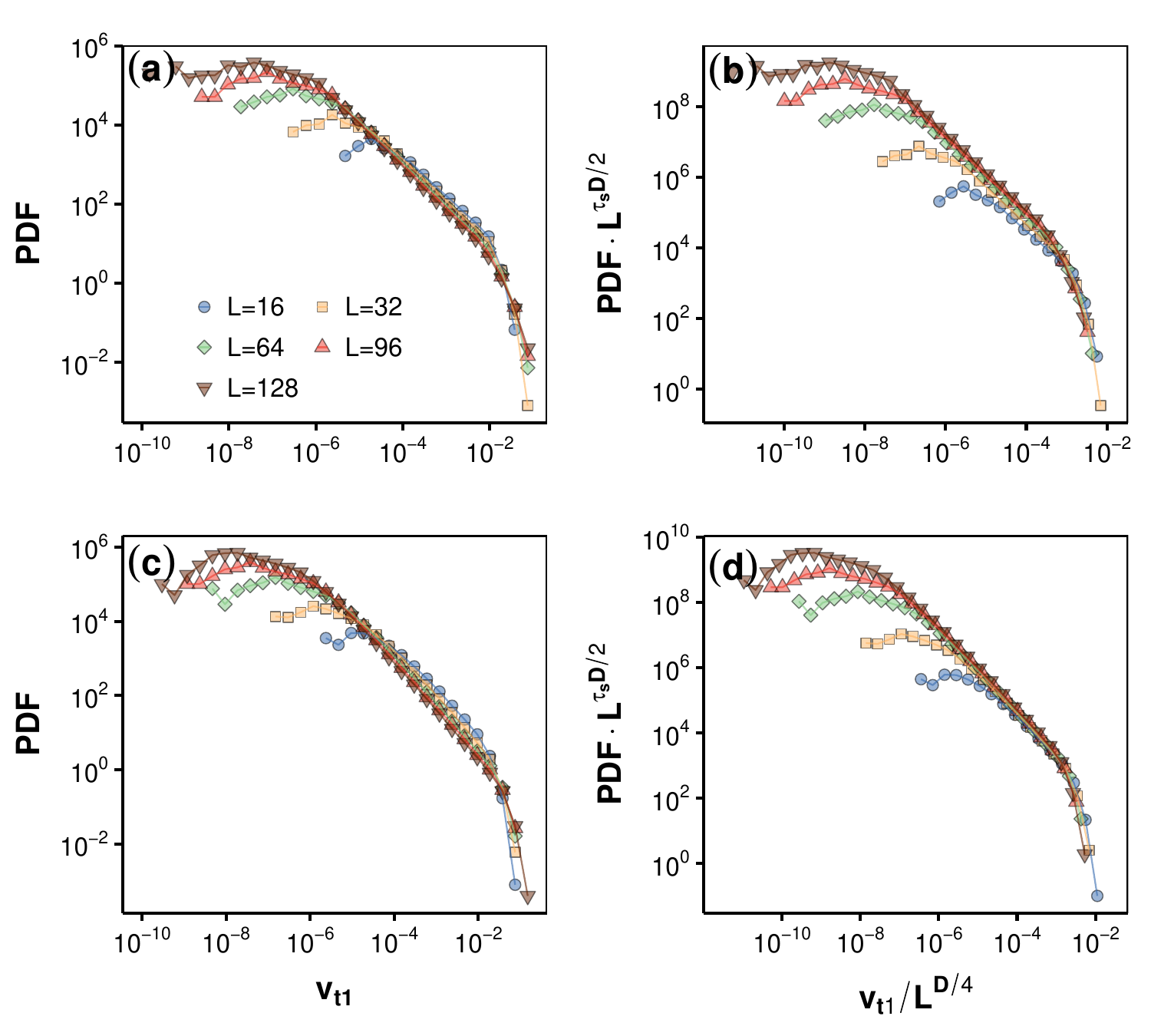}
\caption{\label{fig:5} PDFs of the components of the temporal eigen microstate $\bm{v_1}$ for (a) the BTW model and (c) the Manna model at the critical state with various system sizes. Rescaled PDFs for (b) the BTW model and (d) the Manna model using the critical exponents $\tau_s=1.27$ and $D=2.75$.}
\end{figure}

We then explore the temporal eigen microstate $\bm{v_1}$ at the critical state with varying system sizes. Fig.~\ref{fig:5}(a) and (c) depict PDFs for the BTW model and the Manna model, respectively. Since all components of $\bm{v_1}$ share the same sign (positive or negative), we enforce them to be positive. Three distinct regions emerge in the PDF of $v_{t1}$. For very small $v_{t1}$, the PDF remains unaffected by the increased $v_{t1}$ in Fig.~\ref{fig:5}(a) and (c). In the medium range of $v_{t1}$, the PDF exhibits power-law decay with an exponent ($\sim \tau_s$). Furthermore, for large $v_{t1}$, the PDF rapidly decays, influenced by the system size in Fig.~\ref{fig:5} (a) and (c).

Considering Eq.~(\ref{eq2}) and Eq.~(\ref{eq4}), we can approximate $\langle A_s \rangle \sim \sigma_{0}^{1/2}w_1^{1/2}\tilde{u}_{1}\tilde{v}_{1}$ and $\langle A_{s}^{2} \rangle \sim \sigma_{0}w_1\tilde{u}_1^2$, as the first eigen microstate is dominant. Consequently, we obtain $\langle A_s \rangle^2 / \langle A_{s}^{2} \rangle \sim \tilde{v}_{1}^2$. For both the BTW model and the Manna model, the $k$th moment of the avalanche size follows $\langle A_{s}^{k} \rangle \propto L^{D(1+k-\tau_s)}$ \cite{christensen_complexity_2005}, resulting in $\langle A_s \rangle^2 / \langle A_{s}^{2} \rangle \propto L^{D(1-\tau_s)}$ and $\tilde{v}_{1} \propto L^{D(1-\tau_s)/2}$. To establish the scaling relation, we have
\begin{equation}\label{eq6}
\begin{aligned}
\frac{1}{\sqrt{M}}\tilde{v}_{1}&=\int_{V}v_{t1}PDF(v_{t1};L)dv_{t1}\\  
&=\int_{V}v_{t1}L^{-\tau_s D/2}\mathcal{F}(v_{t1}/L^{\frac{D}{4}})dv_{t1}\\
&=\int_{X}(xL^{\frac{D}{4}})L^{-\tau_s D/2}\mathcal{F}(x)L^{\frac{D}{4}}dx \\ 
&= L^{D(1-\tau_s)/2}\int_{X}x\mathcal{F}(x)dx  \quad   (x=v_{t1}/L^{\frac{D}{4}})\\
&\propto L^{D(1-\tau_s)/2}\;.
\end{aligned}
\end{equation}

Based on Eq.~(\ref{eq6}), we rescale the PDF of $v_{t1}$ with the critical exponents $\tau_s=1.27$ and $D=2.75$ in Fig.~\ref{fig:5}(b) and (d) for the BTW model and the Manna model, respectively. For the medium and large range of $v_{t1}$, the curves collapse across different system sizes. However, scaling breaks for very small $v_{t1}$. We speculate that the inherent fluctuation of the eigenvector related to the ratio $N/M$ \cite{bun2017cleaning} rather than the critical behavior of the system influences the very small $v_{t1}$.

\begin{figure}
\centering
\includegraphics[scale=0.34]{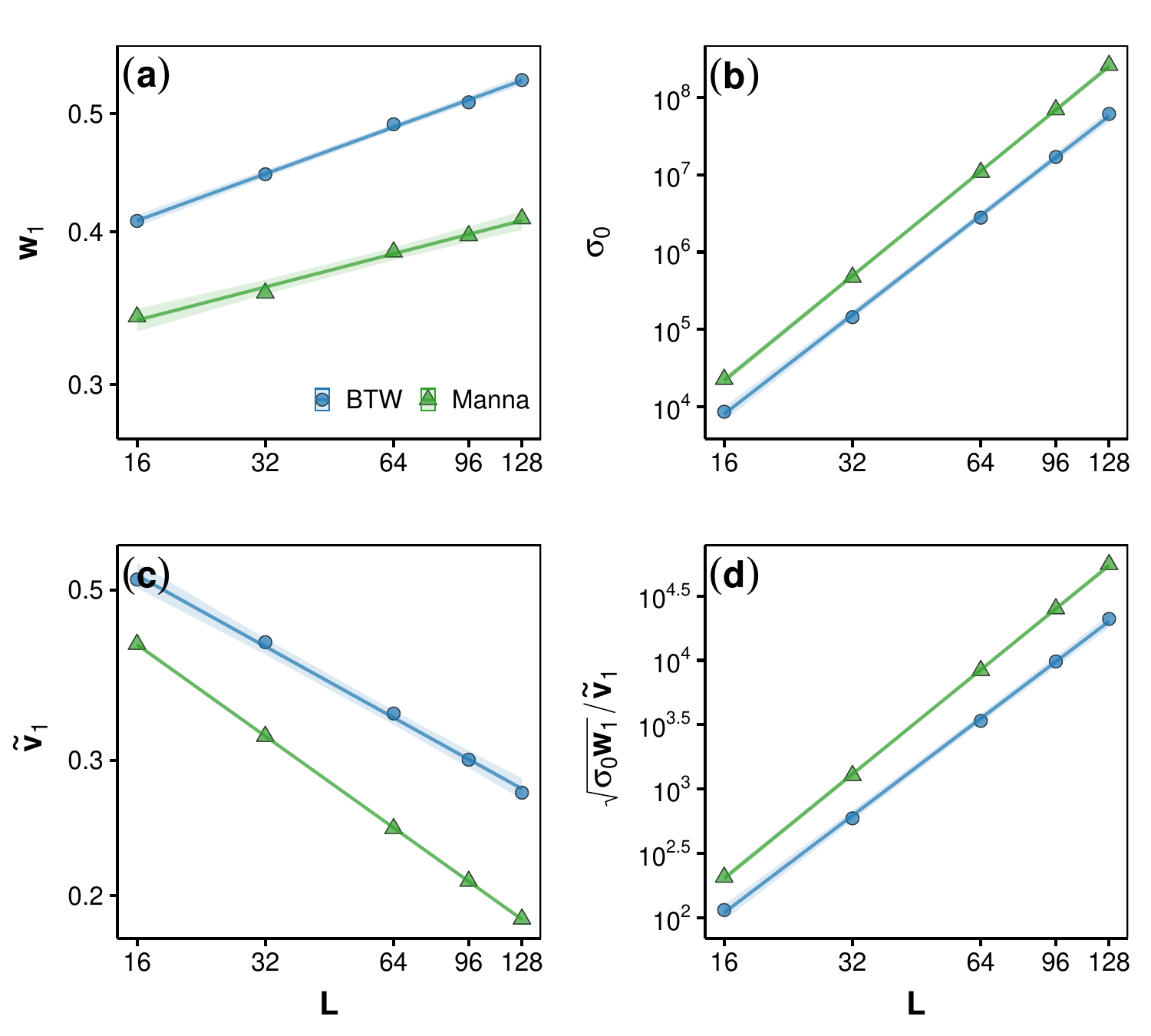}
\caption{\label{fig:6} Log$_{10}$-log$_{10}$ plot of (a) $w_1$, (b) $\sigma_0$, (c) $\tilde{v}_{1}$, and (d) $\sqrt{\sigma_0 w_1}/\tilde{v}_{1}$ as functions of the system size $L$ for both the BTW model and the Manna model at the critical state. The fitted lines, along with confidence intervals, are presented.}
\end{figure}

Next, we study the impact of system size on critical state parameters. The dependence of weight $w_1$, $\sigma_0$, and $\tilde{v}_{1}$ on system size is illustrated in Fig. \ref{fig:6}. We observe $w_1 \propto L^\alpha$ and $\sigma_0 \propto L^\beta$ in Fig.~\ref{fig:6}(a) and (b), where $\alpha$ and $\beta$ are critical exponents detailed in Table~\ref{tab:1}. Additionally, Fig.~\ref{fig:6}(c) demonstrates $\tilde{v}_{1} \propto L^{D(1-\tau_s)/2}$ with $D(1-\tau_s)/2=-0.31\pm0.01$ and $-0.39\pm0.01$ for the BTW model and the Manna model, respectively.

\begin{table}
\caption{\label{tab:1}%
Estimated critical exponents based on the first eigen microstate for both the BTW model and the Manna model at the critical state.   
}
\begin{ruledtabular}
\begin{tabular}{ccccc}
&
$\alpha$ &
$\beta$  &
\multicolumn{1}{c}{$D$}&
$\tau_s$\\
\colrule
\textrm{BTW} & 0.12$\pm$0.01 &  4.27$\pm$0.05 &  2.50$\pm$0.03  &  1.25$\pm$0.05\\
\textrm{Manna} & 0.09$\pm$0.01 &  4.50$\pm$0.03 & 2.69$\pm$0.02 &  1.29$\pm$0.03\\
\end{tabular}
\end{ruledtabular}
\end{table}

By Eq.~(\ref{eq2}) and Eq.~(\ref{eq4}), we find $\langle A_{s}^{2} \rangle / \langle A_{s} \rangle \sim \sqrt{\sigma_0 w_1}\tilde{u}_{1}/\tilde{v}_{1}$. Since $\tilde{u}_{1}$ remains constant with system size,
\begin{equation}\label{eq7}
\sqrt{\sigma_0 w_1}/\tilde{v}_{1} \propto L^D\;.
\end{equation}
Fig.~\ref{fig:6}(d) depicts the outcome, with the estimated critical exponent $D$ presented in Table~\ref{tab:1}. We also estimate the critical exponent $\tau_s$ based on the results of Fig.~\ref{fig:5}(c) and (d), as indicated in Table~\ref{tab:1}. The estimated critical exponents exhibit slight variations between the BTW model and the Manna model at the critical state. Notably, the estimated $\tau_s$ derived from the first eigen microstate is closer to 1.27 than the estimated value from the PDF of avalanche size without corrections.


\begin{figure}
\centering
\includegraphics[scale=0.34]{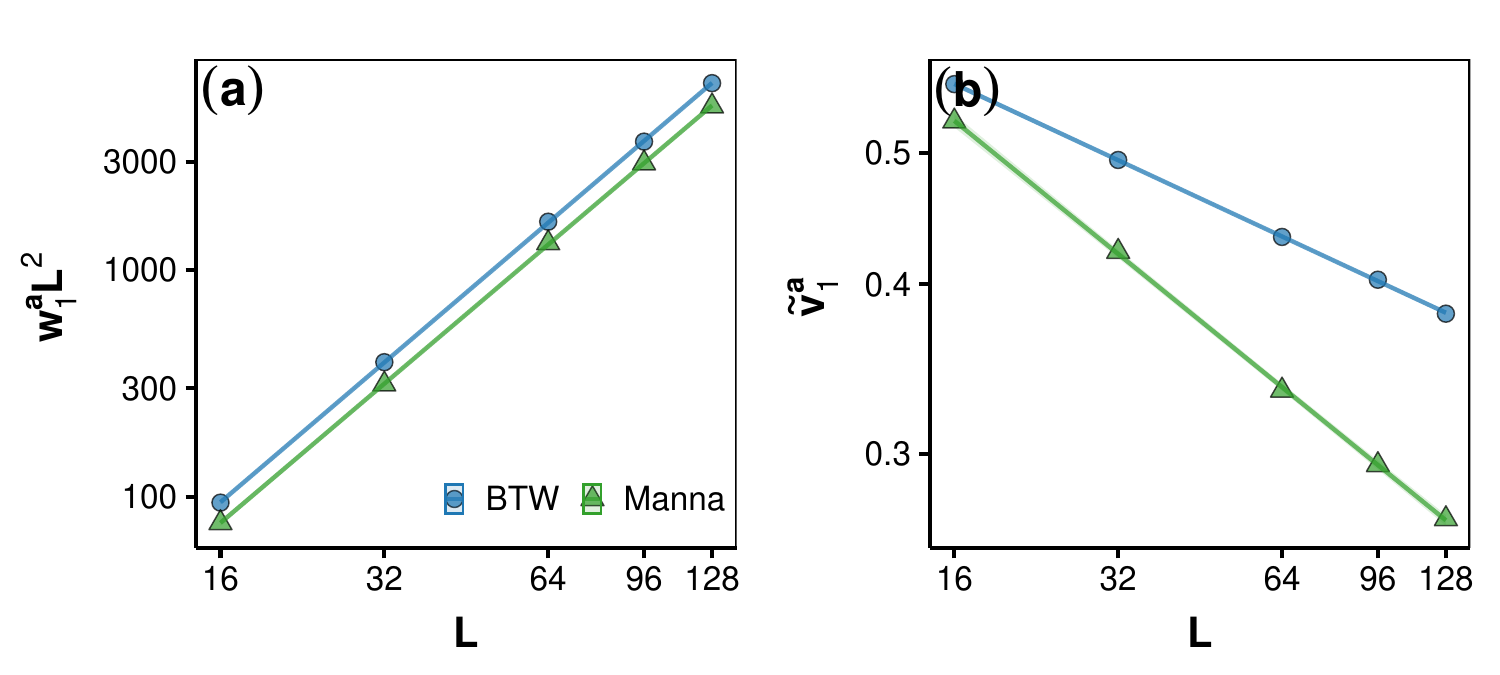}
\caption{\label{fig:7} Log$_{10}$-log$_{10}$ plot of (a) $w^a_1L^2$ and (b) $\tilde{v}^a_{1}$ as functions of the system size $L$ for both the BTW model and the Manna model at the critical state. The fitted lines, along with confidence intervals, are presented.}
\end{figure}

Alternatively, we can define the ensemble matrix as $\bm{S^a}$ with the element $s^a_{it}$ representing the toppling area (multiple topples at the same site will only be counted once during a time step). Similarly, we obtain the eigen ensemble and eigen microstates for the toppling area. Thus, the critical exponents $D_a$ and $\tau_a$ are estimated based on the first eigen microstate. For the toppling area, there is the same relation $\langle A_{a}^{k} \rangle \propto L^{D_{a}(1+k-\tau_a)}$. Regarding the parameter $C^{a}_{0}$, we have $C^{a}_{0}=\sum_{t=1}^{M}\sum_{i=1}^{N}(s^a_{it})^2=\frac{1}{M}\langle A_{a} \rangle$ ($s^{a}_{it}$ can only be 1 or 0), implying that $\sigma^{a}_{0} \propto L^{D_{a}(2-\tau_a)+2}$. Therefore, we deduce the relation $w^{a}_{1}L^2\propto L^{D_a}$ from Eq.~(\ref{eq7}).

Fig.~\ref{fig:7}(a) displays the results. The critical exponents $D_a=2.05\pm0.01$ and $2.04\pm0.01$ are estimated for the BTW model and the Manna model, respectively, both closely approaching $2$ since avalanches in two dimensions are compact \cite{tebaldi_multifractal_1999}. Thus, the critical exponent $\gamma_{sa}=D/D_a$ between the toppling number and area is $1.22\pm0.04$ and $1.32\pm0.03$ for the BTW model and the Manna model, respectively. Previous studies suggested that both $\gamma_{sa}$ for the two models are around $1.35$ or even smaller for BTW \cite{chessa_universality_1999,ben-hur_universality_1996}. Based on the dependence of $\tilde{v}^a_{1}$ on the system size in Fig. \ref{fig:7}(a), we can estimate the critical exponent $\tau_a=1.18\pm0.01$ and $1.32\pm0.01$ for the BTW model and the Manna model, respectively. The obtained exponent is smaller for the BTW model, with its exact value suggested to be $4/3$ \cite{lubeck_numerical_1997}.

\section{\label{sec4}Conclusions}

In conclusion, our investigation into the eigen microstate of SOC through the study of sandpile models, particularly the BTW and Manna models, has revealed intriguing findings. We observed the emergence of dominant eigen microstates associated with amplified weights as these systems transitioned from an absorbing state to a critical state. This behavior mirrors the phase transition phenomena observed in other critical systems, such as the Ising model. Notably, the non-dominant weights, specifically $w_2$ and $w_3,$ unveil themselves as valuable indicators for detecting early warning signals in SOC systems.

Spatial analyses of normalized eigen microstates divulge avalanche characteristics across various spatial scales, from the entire system down to half-scale or smaller. The examination of PDFs for the components of the temporal eigen microstate $\bm{v_1}$ imparts critical behavioral insights. The establishment of scaling relations, coupled with rescaled PDFs demonstrating collapse for medium and large $v_{t1}$ values.

Moreover, we introduced an analysis of finite size effects on the first eigen microstate to estimate critical exponents for both the BTW and Manna models at the critical state i.e., $\sqrt{\sigma_0 w_1}/\tilde{v}_{1} \propto L^D$ and $\tilde{v}_{1} \propto L^{D(1-\tau_s)/2}$. Our investigation extended further to the toppling area, where the critical exponent $\gamma_{sa}=D/D_a$ between toppling number and area revealed values of $1.22\pm0.04$ and $1.32\pm0.03$ for the BTW and Manna models, respectively. The finite size effects of avalanches are predominantly governed by the first eigen microstate, providing more accurate critical exponents associated with the system size.

\begin{acknowledgments}
We thank the financial support by the National Natural Science Foundation of China (Grant No.
12305044 and 12135003) and the National Key Research and Development Program of China
(Grant No. 2023YFE0109000).
\end{acknowledgments}




%

\end{document}